\documentstyle[epsfig, twocolumn]{esapub}

\newcommand{\arcmin}{\hbox{$^\prime$}}                  
\newcommand{\arcsec}{\hbox{$^{\prime\prime}$}}          
\newcommand{\farcs}{\hbox{$.\!\!^{\prime\prime}$}}      

\newcommand{\lesssim}{\mathrel{\hbox{\rlap{\hbox{\lower4pt\hbox{$\sim$}}}\hbox{$<$}}}}
\newcommand{\gtrsim}{\mathrel{\hbox{\rlap{\hbox{\lower4pt\hbox{$\sim$}}}\hbox{$>$}}}}

\newcommand{\eg}{e.g.}                  
\newcommand{\ie}{i.e.}                  
\newcommand{\etal}{et~al.}              

\begin{document}

\setlength{\parindent}{0pt}
\setlength{\parskip}{10pt plus 1pt minus 1pt}
\setlength{\hoffset}{-1.5truecm}
\setlength{\textwidth}{17.1 true cm}
\setlength{\columnsep}{1truecm}
\setlength{\columnseprule}{0pt}
\setlength{\headheight}{12pt}
\setlength{\headsep}{20pt}
\pagestyle{esapubheadings}

\title{\bf STELLAR POPULATIONS BEYOND THE LOCAL GROUP WITH THE {\sl NGST}}

\author{\bf S.~Holland$^1$, B.~Thomsen$^2$ \vspace{2 mm} \\
$^1$Institut for Fysik of Astronomi, Aarhus Universitet, 8000 {\AA}rhus C, Denmark, \\
    tlf $+$45 8942 3601, fax $+$45 8612 0740, e-mail {\tt holland@obs.aau.dk} \\
$^2$Institut for Fysik of Astronomi, Aarhus Universitet, 8000 {\AA}rhus C, Denmark, \\
    tlf $+$45 8942 3617, fax $+$45 8612 0740, e-mail {\tt bt@obs.aau.dk}}

\maketitle


\begin{abstract}

	We present simulated $J$- and $K$-band observations of stars
in the Virgo and Coma clusters of galaxies using the proposed {\sl
Next Generation Space Telescope\/} with a Near-Infrared Camera, and
discuss some of the scientific results that might be obtained.  The
proposed telescope will be able to resolve the brightest $\sim 3$
magnitudes of the red giant branches in the halos of galaxies in the
Virgo Cluster and may be able to resolve stars at the tip of the red
giant branch in the Coma Cluster.  The simulations show that the
background light is more important than the size of the telescope's
aperture in determining the limiting magnitude of the observations.
Therefore we recommend that the {\sl Next Generation Space
Telescope\/} be placed in a $1 \times 3$ a.u.\ orbit to minimize
background light.\vspace{5 pt} \\


Key~words: space astronomy; NGST; stellar populations.

\end{abstract}


\section{INTRODUCTION}
\label{SECTION:intro}

        Before the launch of the {\sl Hubble Space Telescope\/} ({\sl
HST\/}) stellar populations could only be studied in detail in the
Galaxy, the Magellanic Clouds, and the globular star clusters (GCs) of
these galaxies.  The main limitation was not the light-gathering
capabilities of ground-based telescopes, but the crowding due to the
small angular sizes of most external galaxies and the effects of
atmospheric seeing.  The superb resolution of the {\sl HST\/}
significantly reduced the apparent crowding of the stellar images
allowing detailed stellar population studies to be made of many of the
Local Group (LG) galaxies and GCs.  The {\sl HST\/} has also been able
to resolve the brightest 1 to 2 magnitudes of stars in other nearby
galaxies such as Centaurus A and galaxies in the Leo Group.  The major
limitation of the {\sl HST\/} for stellar population studies is the
diameter of the primary mirror (2.4 metres).  The larger mirror on the
proposed {\sl Next Generation Space Telescope\/} ({\sl NGST\/}),
coupled with an expected resolution of $\sim 0\farcs05$ to $0\farcs1$,
will make it possible to resolve individual stars at much greater
distances than the {\sl HST\/} can.  The improved resolution and
light-gathering ability, as well as the reduced background 3 a.u.\
from the Sun, will allow the study of stellar populations in galaxies
in the Virgo Cluster and beyond.

	All other factors being equal, a 4-metre {\sl NGST\/} will
receive $\sim 2.8$ times as much light per unit time as the {\sl
HST\/} does while an 8-metre {\sl NGST\/} will receive $\sim 11$ times
as much light per unit time.  This corresponds to an increase in the
limiting magnitude of 1.12 and 2.60 respectively, which will allow the
study of stellar populations in the Virgo Cluster with a level of
detail that is currently only possible in LG spiral and dwarf
galaxies.  The Virgo cluster, however, contains several elliptical
galaxies and a cD galaxy.  Therefore, the {\sl NGST\/} will allow the
study of stars in galaxies with a wide variety of morphological types
and in a wide variety of environments.  Being able to study the fossil
record of star formation in these galaxies will provide valuable clues
to the way in which these galaxies formed and evolved.  Detailed
studies of M31 ({\eg}~\cite{HF96}, \cite{RM95}) and M33
({\eg}~\cite{MR95}) with the {\sl HST\/} have revealed dramatically
different stellar populations from what are found in the Milky Way
Galaxy, which suggests that the three major LG galaxies have had
radically different formation and enrichment histories despite the
three galaxies having similar morphological types and being located in
the same environment.  These effects must be well understood to
support {\sl NGST\/} studies of galaxies at cosmological distances.


\section{SIMULATED OBSERVATIONS}
\label{SECTION:simulated_data}

	We simulated {\sl NGST\/} observations of fields in the Virgo
and Coma clusters of galaxies using three telescope apertures (4-, 6-,
and 8-metre) and two orbits (1 a.u.\ and $1 \times 3$ a.u.).  We used
the properties of the proposed {\sl NGST\/} and Near-Infrared Camera
(NIRCAM) as described in \cite*{S97}.  Table~\ref{TABLE:NIRCAM} lists
the properties we assumed for the NIRCAM\@.

\begin{table}[htb]
\begin{center}
\caption{\em NIRCAM.}
\vspace{0.5 em}
\begin{tabular}{lcc}
\hline \\[-5 pt]
Property & Value & \\
\hline \\[-5 pt]
Dark Current   &    0.02     & ${\rm e}^- \cdot {\rm s}^{-1} \cdot {\rm pixel}^{-1}$ \\
Full Well      &   60,000    & e$^- \cdot {\rm pixel}^{-1}$ \\
Gain           &      6      & ${\rm e}^- \cdot {\rm ADU}^{-1}$ \\
Read-Out Noise &     15      & ${\rm e}^- \cdot {\rm pixel}^{-1}$ \\
PSF FWHM       & $0\farcs08$ & \\
\hline
\end{tabular}
\label{TABLE:NIRCAM}
\end{center}
\end{table}

	For a {\sl NGST\/} located 1 a.u.\ from the Sun we assumed a
background count rate equal to that of the {\sl HST\/}'s NIC2 camera
and scaled it to the NIRCAM's pixel size.  At 3 a.u.\ from the Sun we
scaled the background count rates as indicated by Figure~2.2 of
\cite*{S97}.  Our adopted $J$- and $K$-band background count rates are
listed in Table~\ref{TABLE:background}.  For long exposures the
background is dominated by zodiacal light and instrument glow
(\cite{S97}).  When the {\sl NGST\/} is 3 a.u.\ from the Sun the
amount of zodiacal light is reduced by a factor of $\sim 30$ to 100.
As shown in Section~\ref{SECTION:virgo} the reduction in the infrared
background obtained by moving the {\sl NGST\/} to 3 a.u.\ from the Sun
corresponds to a significant improvement in the {\sl NGST\/}'s ability
to resolve and photometer individual stars in the Virgo Cluster.  The
actual background count rates in Table~\ref{TABLE:background} include
a contribution from instrument glow so the total reduction in the
infrared background when moving from 1 a.u.\ to 3 a.u.\ is
approximately a factor of 10.

\begin{table}
\begin{center}
\caption{\em The $J$- and $K$-band background count rates in ${\rm
e}^- \cdot {\rm s}^{-1}$ per pixel.}
\vspace{0.5 em}
\begin{tabular}{ccc}
\hline \\[-5 pt]
Distance & $B_J$ & $B_K$ \\
\hline \\[-5 pt]
1 a.u. & 0.015 & 2.044 \\
3 a.u. & 0.002 & 0.204 \\
\hline
\end{tabular}
\label{TABLE:background}
\end{center}
\end{table}

	It is difficult to make detailed simulations of observations
with a space telescope that has not been built and whose orbit has not
been determined.  The following simulations are intended to show the
relative capabilities of several different {\sl NGST\/} apertures and
orbits, and to compare the capabilities of the {\sl NGST\/} with those
of the {\sl HST}.  The simulations ignore cosmic rays (which are
expected to be negligible for multiple exposure times of $\lesssim
1000$ seconds each), dust in the target galaxies (which should be
negligible for observations in elliptical galaxies and in the halos of
spiral galaxies), flat-fielding errors in the NIRCAM, and
contamination due to foreground and background objects.  All of these
will add small amounts of scatter to the observed CMDs and act to
slightly reduce the limiting magnitude of the data.  However,
observations can be planned to minimize contamination from foreground
and background objects, and image classification software such as {\sc
Sextractor} is very efficient at separating background galaxies from
stars ({\eg}~\cite{RG98}) and thus removing the systematic bias
background objects can introduce into a CMD\@.  Sophisticated data
reduction and calibration techniques, such as the methods used by
\cite*{RF97} to obtain deep CMDs of white dwarfs in the M4, will
further reduce the amount of photometric scatter, improve the accuracy
of the photometry, and allow fainter stars to be recovered and
photometered.


\subsection{The Virgo Cluster}
\label{SECTION:virgo}

	We constructed $J$- and $K$-band images of an artificial star
field at a projected distance of $172\arcsec$ ($\sim 13.3$ kpc) from
the core of M87.  This is twice the effective radius of the galaxy so
the unresolved background light from M87 is small compared to the
contribution from the sky and telescope glow.  We added 115,000
artificial stars based on the colour--magnitude diagram (CMD) and
luminosity function (LF) of the upper red giant branch (RGB) of M13.
Background light, Poisson noise, and read-out noise were added to each
image.  Contamination from cosmic rays is not expected to affect more
than $\sim 5$\% of the pixels for exposure times of $\lesssim 1000$
seconds (\cite{S97}) so we set our exposures to 1000 seconds and
ignored cosmic rays.  We simulated $36 \times 1000$ second exposures
in each of the $J$- and $K$-bands.  This corresponds to one day of
observing time if we assume 2.5 minutes of overhead time per exposure,
or 28 {\sl HST\/} orbits.  Combined $J$- and $K$-band images were
created for each field by taking the mean of all the exposures in each
filter.  The resulting combined images were reduced using the {\sc
DaoPhot II} digital photometry software package (\cite{S87}).
Aperture corrections were computed based on the input and recovered
magnitudes of the brightest star in each field.

\begin{figure}[!ht]
\begin{center}
\centerline{\psfig{file=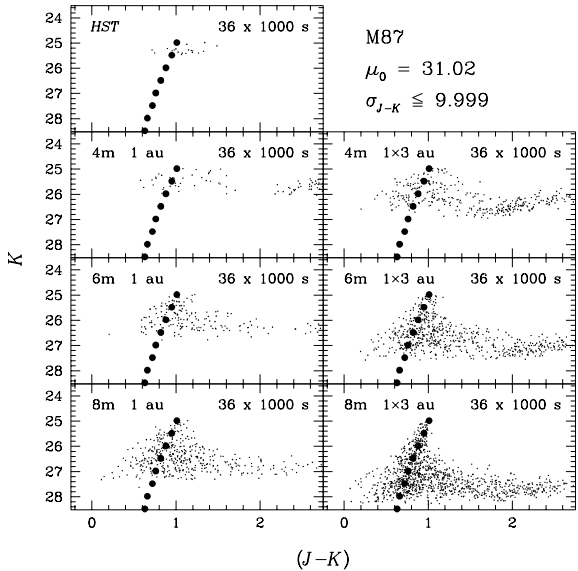,height=8.0cm,width=8.0cm}}
\end{center}
\caption{\em This figure shows all of the recovered stars that appear
in both the $J$- and $K$-band images for our simulated M87 data.  The
dots are the recovered stars while the large solid circles show the
fiducial sequence for the input RGB\@.  The left-hand label indicates
the size and orbit of the telescope while the right-hand label
indicates the number of exposures per filter.}
\label{FIGURE:virgo_cmd_all}
\end{figure}

\begin{figure}[!ht]
\begin{center}
\centerline{\psfig{file=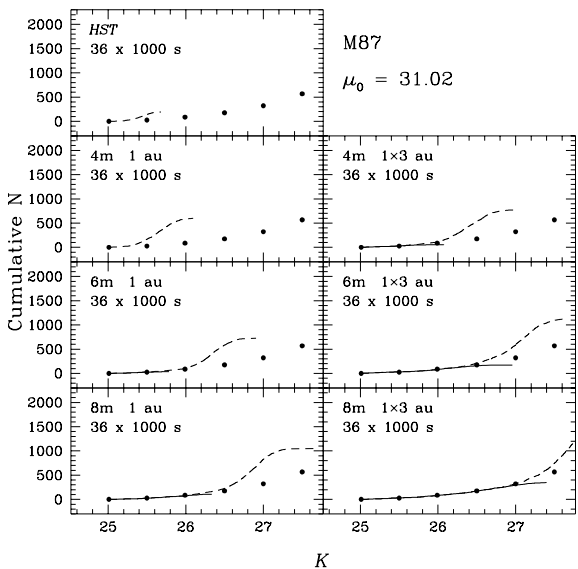,height=8.0cm,width=8.0cm}}
\end{center}
\caption{\em This figure shows the input and recovered $K$-band LFs
for the M87 simulations.  The filled circles show the input LF, the
dashed line shows the LF for all of the detections, and the solid line
shows the LF for those detections with $\sigma_K \le 0.2$
({\ie}~S/N~$\ge 5$).}
\label{FIGURE:virgo_lf}
\end{figure}

	Figure~\ref{FIGURE:virgo_cmd_all} shows the CMDs for the
simulated {\sl HST\/} and {\sl NGST\/} observations of the halo of
M87.  All stars that were recovered in both the $J$- and $K$-band
images are plotted.  Figure~\ref{FIGURE:virgo_lf} shows the $K$-band
LFs for each simulation.  The observed LF only agrees with the input
LF down to some limiting magnitude, which depends on the telescope's
aperture and distance from the Sun.  Fainter than this limiting
magnitude the recovered LF rises more rapidly than the input LF does,
which suggests that many of these detections are not real stars.
Figure~\ref{FIGURE:virgo_lf} suggests that detections with $\sigma
\lesssim 0.2$ (corresponding to $\sigma_{J\!-\!K} \lesssim 0.3$) are
probably real stars.  Therefore we have taken the limiting magnitude
of the simulated observations to be the magnitude at which $\sigma =
0.2$.  This corresponds to a signal-to-noise ratio (S/N) of 5.

\begin{figure}[!ht]
\begin{center}
\centerline{\psfig{file=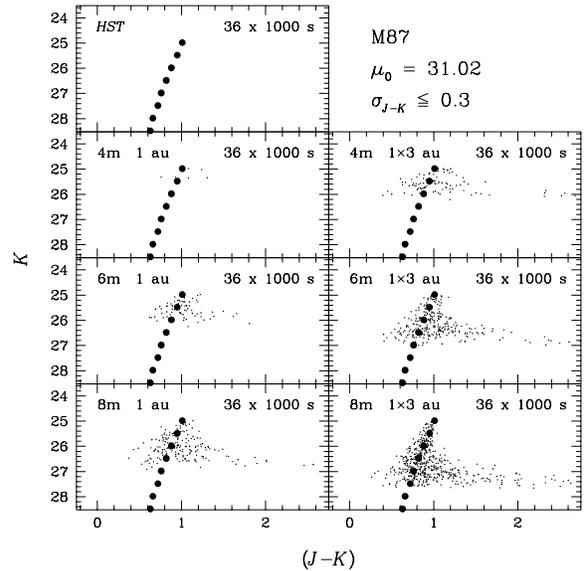,height=8.0cm,width=8.0cm}}
\end{center}
\caption{\em This figure shows only those stars with S/N $\ge 5$ in
each filter (corresponding to stars with $\sigma_{J\!-\!K} \le 0.3$).
As discussed in Section~\ref{SECTION:virgo} most of these objects are
real detections of stars in the halo of M87.}
\label{FIGURE:virgo_cmd_03}
\end{figure}

	Figure~\ref{FIGURE:virgo_cmd_03} shows recovered stars with
S/N $\ge 5$, {\ie}~stars brighter than the limiting magnitude of the
photometry.  Most of the stars in this figure are real stars, not
false detections, although the fraction of false detections will
increase near the limiting magnitude of each CMD\@.
Table~\ref{TABLE:virgo_limit} lists the estimated limiting magnitudes
of each of the simulated M87 observations.  None of the 19 recovered
stars in the simulated {\sl HST\/} observations had S/N $\ge 5$.

\begin{table}
\begin{center}
\caption{\em Estimated limiting magnitudes (where S/N $\ge 5$) for
each of the M87 simulations.}
\vspace{0.5 em}
\begin{tabular}{ccc}
\hline \\[-5 pt]
Telescope & $J_{\lim}$ & $K_{\lim}$ \\
\hline \\[-5 pt]
{\sl HST}            & $\cdots$ & $\cdots$ \\
 4m     1 au         & 26.3 & 25.3 \\
 6m     1 au         & 27.0 & 26.0 \\
 8m     1 au         & 27.6 & 26.6 \\
 4m $1{\times}3$ au  & 27.0 & 26.0 \\
 6m $1{\times}3$ au  & 27.9 & 26.9 \\
 8m $1{\times}3$ au  & 28.6 & 27.6 \\
\hline
\end{tabular}
\label{TABLE:virgo_limit}
\end{center}
\end{table}

	These results suggest that a telescope aperture of at least 6
metres will be needed to do stellar population studies in galaxies in
the Virgo cluster if the {\sl NGST\/} is placed in a 1 a.u.\ orbit.
The reduction in the background obtained by using a $1 \times 3$ a.u.\
orbit results in the NIRCAM's limiting magnitude becoming $\sim 1$
magnitude fainter.  This would enable similar stellar population
studies to be undertaken with a 4-metre {\sl NGST\/} but the best
results were obtained from a 6- or 8-metre {\sl NGST\/} in a $1 \times
3$ a.u.\ orbit.  In this scenario it should be possible to resolve the
brightest two or three magnitudes of the RGB in M87 and other Virgo
Cluster galaxies with $\sim 1$ day of observing time.

	Horizontal branch (HB) stars in M87 should be located at $K
\sim 28.5$.  These were not included in the simulation shown since
they were not recovered with $36 \times 1000$ second exposures.  Our
simulations suggest that HB stars may be observable with $\sim 30$
hours of observing in {\sl each\/} of the $J$- and $K$-bands with an
8-meter {\sl NGST\/} in a $1 \times 3$ a.u.\ orbit.


\subsection{The Coma Cluster}
\label{SECTION:coma}

\begin{figure}[!ht]
\begin{center}
\centerline{\psfig{file=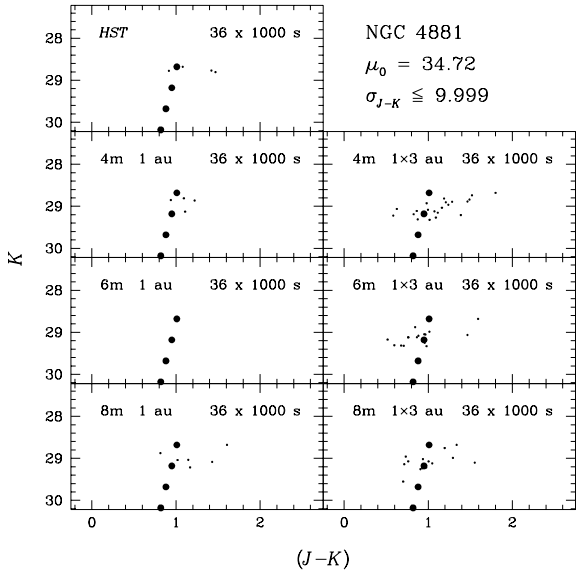,height=8.0cm,width=8.0cm}}
\end{center}
\caption{\em This figure shows all of the recovered stars that appear
in both the $J$- and $K$-band images for our simulated NGC 4881 data.
None of these stars have S/N greater than $\sim 2.5$.}
\label{FIGURE:coma_cmd_all}
\end{figure}

\begin{figure}[!ht]
\begin{center}
\centerline{\psfig{file=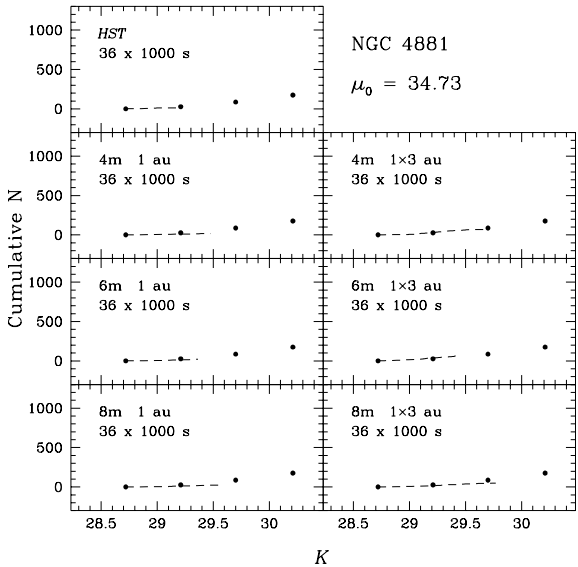,height=8.0cm,width=8.0cm}}
\end{center}
\caption{\em This figure shows the input LF (filled circles) and
recovered LF (dashed line) for the NGC 4881 simulations.  For all
configurations of the\/ {\em NGST} only the stars at the tip of the
RGB are recovered and these stars have S/N $< 5$.}
\label{FIGURE:coma_lf}
\end{figure}

	Recent {\sl HST\/} WFPC2 observations of the tip of the RGB in
Virgo Cluster galaxies (\cite{HD98}) suggest that it may be possible
to observe the tip of the RGB in the Coma Cluster using the {\sl
NGST\/} under optimum conditions.  We simulated these observations in
the manner described in Section~\ref{SECTION:virgo} We constructed an
artificial star field in the halo of NGC 4881 at a projected distance
of $34\arcsec$ ($\sim 67$ kpc, or $\sim 2$ effective radii) from the
core of NGC 4881, and reduced the images in exactly the same manner as
was done for the simulated M87 observations.  The resulting CMDs are
shown in Figure~\ref{FIGURE:coma_cmd_all}.  All stars that were
recovered in both the $J$- and $K$-bands are plotted.  The input and
recovered LFs are shown in Figure~\ref{FIGURE:coma_lf}.

	With an 8-metre {\sl NGST\/} in a $1 \times 3$ a.u.\ orbit our
simulations suggest that stars at the tip of the RGB might be
observable in a manner similar to the recent observations of the tip
of the RGB in the Virgo Cluster (\cite{FT98}, \cite{HD98}).  The stars
that were recovered in both the $J$- and $K$-band images had S/N of
less than 5, so it is not certain if any of the recovered stars are
real.  Table~\ref{TABLE:coma_sn} lists the mean observed S/N for the
recovered stars at the tip of the RGB for each simulated NGC 4881
observation.

\begin{table}
\begin{center}
\caption{\em The mean S/N of the stars at the tip of the RGB in the
simulated observations of NGC 4881.}
\vspace{0.5 em}
\begin{tabular}{ccc}
\hline \\[-5 pt]
Telescope & $\overline{{\rm S/N}_J}$ & $\overline{{\rm S/N}_K}$ \\
\hline \\[-5 pt]
{\sl HST}           & 2.1 & 1.8 \\
 4m     1 au        & 2.2 & 2.3 \\
 6m     1 au        & 1.6 & 1.7 \\
 8m     1 au        & 2.3 & 2.2 \\
 4m $1{\times}3$ au & 2.1 & 2.1 \\
 6m $1{\times}3$ au & 2.5 & 2.3 \\
 8m $1{\times}3$ au  & 2.2 & 2.2 \\
\hline
\end{tabular}
\label{TABLE:coma_sn}
\end{center}
\end{table}


\section{SOME SCIENTIFIC OBJECTIVES}
\label{SECTION:science}

	Two of the primary goals of the {\sl NGST\/} are (1) to study
the formation of galaxies and (2) to study the structure and dynamics
of galaxies at redshifts of $z \sim 2$.  Galaxies at large redshifts
can only be studied through integrated colours and spectra.  The
detailed physical properties of these galaxies are then deduced
through these integrated properties.  In order to understand the
relationship between the integrated properties of high-redshift
galaxies and their stellar populations it is necessary to understand
how the stellar content of a galaxy affects the observed integrated
light and spectra.

	Some of the results of studying stellar populations in a large
number of galaxies of different Hubble types and in different
environments are discussed below.


\subsection{Age}
\label{SECTION:age}

	If a Virgo Cluster galaxy is currently undergoing star
formation the upper main sequence will be resolved with the {\sl
NGST}.  If the main-sequence turn-off can be located, the age of the
stars can then be determined with a high degree of precision.  An
intermediate-aged population can be identified by the presence of
super-luminous asymptotic giant branch (AGB) stars.  Photometric
scatter in the RGB will make distinguishing between RGB and AGB stars
difficult for some of the proposed {\sl NGST\/} designs.  However, if
the {\sl NGST\/} is placed in a $1 \times 3$ a.u.\ orbit and has a 6-
or 8-metre aperture detailed studies will be possible.  Studies of
halo and bulge fields in M31 (\cite{HF96}, \cite{RM95}) as well as
fields in M33 (\cite{MR95}) show that it is possible to use the lack
of super-luminous AGB stars to constrain the age distribution.  The
uncertainties in the photometry obtained in those studies were similar
to the photometric scatter in our simulated {\sl NGST\/} observations
of the RGB of M87.  The record of star formation contained in the
brightest few magnitudes of a CMD can be combined with dynamical
information on other Virgo Cluster galaxies to study the relationship
between close encounters between galaxies and star formation in a
large number of galaxies of different morphological types.


\subsection{Metallicity}
\label{SECTION:metallicity}

	There is a relationship between the shape of the upper RGB and
metallicity (\cite{DA90}).  If a stellar population contains a range
of metallicities, the RGB will be broader than would be expected from
the photometric uncertainties alone.  A grid of model RGBs with known
metallicities can be placed over the observed RGB and used to
interpolate a metallicity value for each star.  The broadening of the
derived metallicity distribution due to photometric scatter can be
removed by deconvolving the observed error distribution from the
metallicity distribution.  This technique has been used to determine
the metallicities of several GCs in LG galaxies ({\eg}~\cite{HF97})
and to find the metallicity distribution of stars in the halo of M31
({\eg}~\cite{HF96}).  Our results suggest that the metallicity
distributions of stellar populations in the Virgo Cluster can be
estimated from the curvatures and widths of their RGBs.  The high
resolution and large field of view of the NIRCAM ($4\arcmin \times
4\arcmin$) will make it possible to rapidly map the dependence of
metallicity on position in the halo of a galaxy in the Virgo Cluster.
This may allow the identification of recent mergers by locating areas
of anomalous metallicity in a galaxy's halo.  It can also be used to
trace star formation history as a function of position in a galaxy.

	CMD techniques can be used to determine the metallicity of
stellar populations that are too diffuse to permit spectral
determinations of elemental abundances.  Since CMD techniques work
best in uncrowded fields they are a natural complement to {\sl NGST\/}
spectral observations of the bulges and cores of galaxies.


\subsection{Distance}
\label{SECTION:distance}

	One of the Key Projects for the {\sl HST\/} was to investigate
the extragalactic distance scale using Cepheid variables in galaxies
in the Virgo Cluster ({\eg}~\cite{KS98}).  High-quality CMDs and
colour--colour diagrams of the fields containing Virgo Cepheids will
allow improved estimates of the reddening and metallicity corrections
that need to be applied to Cepheid observations.  Both the tip of the
RGB and the curvature of the upper RGB can be used to determine the
distance to a galaxy if the metallicity is known.  An 8-metre {\sl
NGST\/} in a $1 \times 3$ a.u.\ orbit will be able to resolve HB stars
with $\sim 60$ hours of observing time.  This will provide a new means
of measuring the distance to the Virgo Cluster.

	Figures~\ref{FIGURE:coma_cmd_all}~and~\ref{FIGURE:coma_lf}
show that it may be possible to resolve individual stars at the tip of
the RGB in the Coma cluster of galaxies in a similar manner to what
has been done in the Virgo Cluster using the WFPC2 aboard the {\sl
HST}.  This would provide a powerful new method of determining the
distance to the Coma Cluster.  Since this cluster is well beyond the
influence of the Virgocentric infall it is ideal for determining the
value of the Hubble Parameter.


\section{CONCLUSIONS}
\label{SECTION:conclusions}

	All of the proposed designs for the {\sl NGST\/} will allow
colour--magnitude studies of stars in the Virgo Cluster to be
performed, although better results are obtained with larger apertures
and greater distances from the Sun.  Reducing the amount of background
light is more important than increasing the telescope's aperture for
resolving stars in distant galaxies.  This is because reducing the
background results in a reduction in the amount of Poisson noise.
This leads to an increase in the signal-to-noise ratio for faint stars
which makes them easier to detect and photometer.  The same is true
for resolving stars in Coma Cluster galaxies.

	Our results suggest that the {\sl NGST\/} will be able to
study resolved stellar populations in Virgo Cluster galaxies at
similar levels of detail similar to studies of LG galaxies using large
ground-based telescopes.  If an 8-metre {\sl NGST\/} in a $1 \times 3$
a.u.\ orbit observes a Virgo Cluster galaxy for only $\sim$ 7 days the
resulting observation will be comparable to the best existing {\sl
HST\/} observations of stars in the halo of M31.


\section*{ACKNOWLEDGMENTS}

	This work was supported by a grant to the Danish Centre for
Astrophysics with the {\sl HST\/} from the Danish Research Councils.



\end{document}